\documentclass[twocolumn,aps,pre,superscriptaddress]{revtex4-2}
\usepackage[latin9]{inputenc}
\usepackage{verbatim}
\usepackage{amsmath}
\usepackage{amssymb}
\usepackage{graphicx}

\makeatletter
\usepackage{eqnarray}\usepackage{amsthm}
\usepackage{fancyhdr}
\usepackage{subfigure}\usepackage{epsfig}
\usepackage{epsfig}
\usepackage{bbm}
\usepackage{subfigure}
\usepackage{bm}
\usepackage{float}
\usepackage{xcolor}

\newcommand{\I}{\boldsymbol{1}}

\newcommand{\rot}{\mathcal{R}}
\newcommand{\order}{\mathcal{O}}

\newcommand{\ssm}{s_{-}}
\newcommand{\ssp}{s_{+}}
\newcommand{\ccm}{c_{-}}
\newcommand{\ccp}{c_{+}}
\newcommand{\ccpm}{c_{\pm}}
\newcommand{\sspm}{s_{\pm}}

\newcommand{\xxi}{\boldsymbol{\xi}}
\newcommand{\xxia}{\xxi_1}
\newcommand{\xxib}{\xxi_2}
\newcommand{\eeta}{\boldsymbol{\eta}}

\newcommand{\ssig}{\boldsymbol{\sigma}}
\newcommand{\pp}{\boldsymbol{p}}

\newcommand{\atan}{\text{atan}}

\newcommand{\rr}{\boldsymbol{r}}
\newcommand{\rra}{\boldsymbol{r}_1 }
\newcommand{\rrb}{\boldsymbol{r}_2 }
\newcommand{\RR}{\boldsymbol{R}}
\newcommand{\nn}{\boldsymbol{n}}
\newcommand{\nna}{\boldsymbol{n}_1}
\newcommand{\nnb}{\boldsymbol{n}_2}

\newcommand{\JJ}{\boldsymbol{J}}

\newcommand{\dt}{\partial_t}

\newcommand{\dx}{\partial_{x}}

\newcommand{\nabR}{\nabla}

\newcommand{\nabRd}{\tilde{\nabla}}

\newcommand{\FF}{\boldsymbol{F}}

\newcommand{\fs}{f_s}

\newcommand{\fa}{\fs(\rra)}
\newcommand{\fb}{\fs(\rrb)}

\newcommand{\n}{n}
\newcommand{\nx}{n_x}
\newcommand{\ny}{n_y}

\newcommand{\sn}[1]{{\color{red}{\bf #1}}}
\newcommand{\cmnt}[1]{{\color{black}{ #1}}}

\makeatother

\begin{document}

\title{Active colloidal molecules in activity gradients}
\author{Hidde D. Vuijk}
\affiliation{Leibniz-Institut f\"ur Polymerforschung Dresden, Institut Theory der Polymere, 01069 Dresden, Germany}
\author{Sophie Klempahn}
\affiliation{Leibniz-Institut f\"ur Polymerforschung Dresden, Institut Theory der Polymere, 01069 Dresden, Germany}
\author{Holger Merlitz}
\affiliation{Leibniz-Institut f\"ur Polymerforschung Dresden, Institut Theory der Polymere, 01069 Dresden, Germany}
\affiliation{School of Physical Science and Technology, Xiamen University, Xiamen 361005, PR China}
\author{Jens-Uwe Sommer}
\affiliation{Leibniz-Institut f\"ur Polymerforschung Dresden, Institut Theory der Polymere, 01069 Dresden, Germany}
\affiliation{Technische Universit\"at Dresden, Institut f\"ur Theoretische Physik, 01069 Dresden, Germany}
\author{Abhinav Sharma}
\affiliation{Leibniz-Institut f\"ur Polymerforschung Dresden, Institut Theory der Polymere, 01069 Dresden, Germany}
\affiliation{Technische Universit\"at Dresden, Institut f\"ur Theoretische Physik, 01069 Dresden, Germany}

\begin{abstract}
We consider a rigid assembly of two active Brownian particles, forming an active colloidal dimer, in a gradient of activity.
We show analytically that depending on the relative orientation of the two particles the active dimer accumulates in regions of either high or low activity,
corresponding to, respectively, chemotaxis and antichemotaxis.
Certain active dimers show both chemotactic and antichemotactic behavior, depending on the strength of the activity.
Our coarse-grained Fokker-Planck approach yields an effective potential,
which we use to construct a nonequilibrium phase diagram that classifies the dimers according to their tactic behavior.
Moreover, we show that for certain dimers a higher persistence of the motion is achieved similar to the effect of a steering wheel in macroscopic devices.
This work could be useful for designing autonomous active colloidal structures which adjust their motion depending on the local activity gradients.
\end{abstract}
\maketitle

\section{introduction}
Active matter has a wide range of applications \citep{jurado-sanchez2017perspectives}: 
material science \citep{needleman2017active},
environmental science (e.g. clean up of pollutants \citep{parmar2018micro,jurado-sanchez2018micromotors,zarei2018selfpropelled}),  transport of cargo
\citep{ma2015catalytic,baraban2012transport,merlitz2017directional,vuijk2021chemotaxis},
and biomedical science 
\citep{wang2012nano,mathesh2020enzyme,wang2019biocompatibility}
(e.g. drug delivery \citep{mathesh2020supramolecular,reinisova2019micro,kim2018artificial,park2017multifunctional,deavila2017micromotorenabled,xuan2014self,qiu2015magnetic,gao2014synthetic}).
For many applications it is important to steer the active particles towards the correct target zone.
Steering of active particles has been realized experimentally by feedback mechanisms,
where the state of the active particle (position and orientation) are measured and accordingly the external stimuli are modified \citep{mano2017optimal,qian2013harnessing}.
However, since it is not always be possible to externally measure the state and tune the behaviour of active particles, an autonomous approach is desirable where an active particle \emph{senses} the local environment and adjusts its behaviour accordingly. 

A way to control the behaviour of active particles is by subjecting them to spatial external fields.
For example a space-dependent friction \citep{liebchen2018viscotaxis} or a space-dependent swim force \citep{sharma2017brownian,stenhammar2016lightinduced,soker2021how,auschra2021polarizationdensity,caprini2022active,caprini2022dynamics}.
Here we focus on the latter.
It is well known that objects accumulate where they are less agitated.
For active particles this means that they accumulate where the swim force is small
\citep{sharma2017brownian,moran2021chemokinesisdriven,schnitzer1993theory}.
Active particles with a space dependent swim force give rise to a wide variety of behaviour that has consequences for
their tactic properties 
\citep{schnitzer1993theory,lozano2016phototaxis, vuijk2018pseudochemotaxis,ghosh2015pseudochemotactic,merlitz2020pseudochemotaxis},
search strategies \citep{kromer2020chemokinetic,kromer2021composite},
trapping \citep{jahanshahi2020realization},
and, when the swim force is both space and time dependent, can be used to induce a flux
\citep{merlitz2018linear,lozano2019propagating,lozano2019diffusing,geiseler2016chemotaxis,geiseler2017selfpolarizing}.

\begin{figure}
\centering
\includegraphics[width=.45\linewidth]{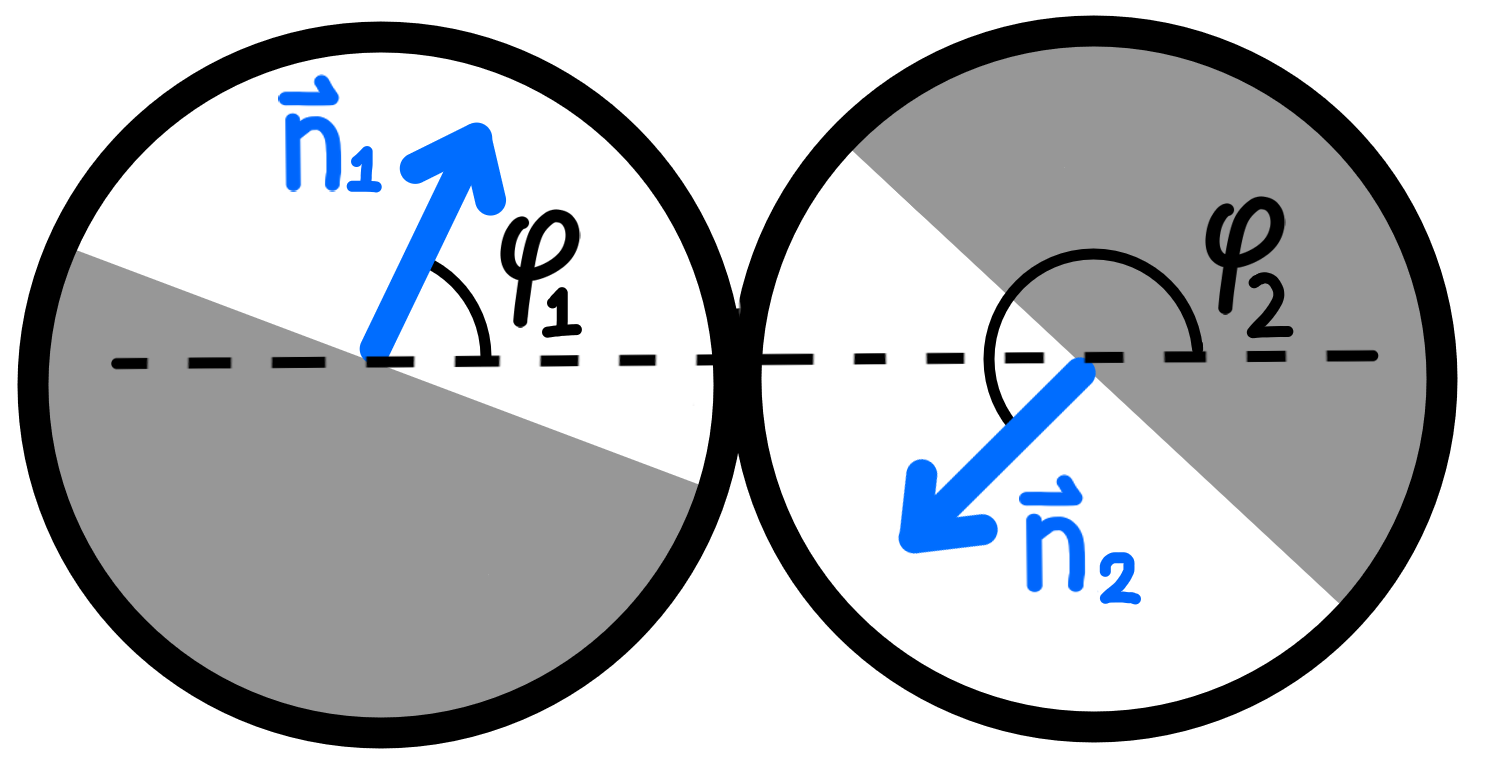}
\caption{A sketch of an active colloidal dimer consisting of two active Janus particles.
The orientation vectors $\nna$ and $\nnb$  of the active particles are shown in blue.
The angles $\phi_1$ and $\phi_2$ are the angles between $\nna$ and $\nnb$ and  the vector connecting the centers of the particle.
}
\label{fig:dimer}
\end{figure}

Colloidal sized active Brownian particles (APBs) can be assembled into active colloidal molecules \citep{lowen2018active},
for example, dimers and tadpole shaped particles \citep{ebbens2010selfassembled,nourhani2016spiral,johnson2017dynamic},
active polymers \citep{yan2016reconfiguring},
or more complex structures \citep{mallory2017selfassembly,schmidt2019lightcontrolled}.
From a theoretical perspective, active particles connected in a chain to form polymers, have recently received much attention 
\cmnt{\citep{vuijk2021chemotaxis,winkler2017active,mousavi2019active,winkler2020physics,martin-gomez2019active,eisenstecken2022path,smrek2020active,chubak2020emergence,smrek2017small,prathyusha2022emergent,bianco2018globulelike}}.
In contrast to previous work \citep{vuijk2021chemotaxis}, here we consider an active dimer where the orientation of the active particles that constitute the dimer are fixed with respect to the bond vector (see Fig. \ref{fig:dimer}), which corresponds to the experimental systems in refs. \citep{ebbens2010selfassembled,nourhani2016spiral,johnson2017dynamic}.
In particular, we study the behaviour of active colloidal dimers with a space dependent swim force,
and how the orientation of the active particles relative to the bond vector affects the dimer's behaviour,
as proposed in Ref.  \cite{attard2012design}.
With the recent advances in fabrication techniques, colloidal particles can now be assembled into desired structures \citep{chen2011directed,bianchi2011patchy,sacanna2010lock,glotzer2007anisotropy,zhang2016directed,popescu2020chemically}.
Since structure determines the functionality of the active dimer
our study could be important for the design of active matter for environmental and medical applications where,
generally, one has little or no control over the external gradients \cite{golestanian2007designing}.

\section{Model}
We consider a two dimensional model consisting of two ABPs \citep{bechinger2016active} attached to each other forming an active colloidal dimer,
see Fig. \ref{fig:dimer}.
The motion of the dimer is governed by the following Stochastic Differential Equations (SDEs):
\begin{align}
\dt \rra &=
  ~~  4 \FF + 4 \fa \nna + 2 \xxia, \label{eomr1d}
\\
\dt \rrb &=
 -   4 \FF + 4 \fb \nnb + 2 \xxib, \label{eomr2d}
\end{align}
where 
$\rra$ ($\rrb$) is the position of particle 1 (2),
the vector $\xxia$ and $\xxib$ are random Gaussian vectors with
$\left< \xxi_1(t) \right> = \left< \xxi_2(t) \right> = 0$,
$\left< \xxi_1(t) \xxi_1(t') \right> = \left< \xxi_2(t) \xxi_2(t') \right> =  \I \delta(t - t') $,
$\fs(\rr)$ is the active force at position $\rr$, and
$\nn_1$ ($\nn_2$) is the direction of the active force on particle one (two).
The force $\FF$ holds the two particles together.
In the following we take this force to be strong enough to keep the two particles at 
a fixed distance $l$ from each other.
Furthermore, this force fixes the orientation of the two ABPs relative to the bond vector.
The unit of length and time are such that $l=1$ and the diffusion constant of  the center-of-mass coordinate of the dimer is unity. 
The unit of force is $2T/l$, where $T$ is the temperature in units such that the Boltzmann constant is unity.
Note that in contrast to other theoretical studies \citep{caprini2020hidden,sharma2017brownian,farage2015effective,sharma2017escape}, the rotational diffusion constant is not a free parameter,
but comes from the translational diffusion of the two particles.
Note that, in order to keep our analysis general, we do not take into account the torque on the two active particles due to the activity gradient because this depends on the specific self-propulsion mechanism
\citep{golestanian2009anomalous}.
However, this torque can be included in the analysis presented here.
We ignore the hydrodynamic interaction between the two particles, and their
effect on the self-propulsion
\citep{popescu2011pulling,popescu2018effective,reigh2015catalytic,reigh2018diffusiophoretically}.

Because the distance between the two particles is constant, the two translational degrees of freedom of the two particles can be transformed to the center-of-mass
coordinate of the dimer $\RR = (\rra + \rrb)/2$ and $\theta$, the angle between the bond vector
$\nn = \rra - \rrb =  \left( \cos \theta, \sin \theta \right)$ and the x-axis.
We call the bond vector $\nn$ the orientation of the dimer.
The corresponing SDEs are
\begin{align}
\dt \RR &=
    2 \left[ \fa \nna + \fb \nnb \right] + \sqrt{2} \xxi,
    \label{eomRd}
    \\
\dt \theta &=
    -4 \nn {\cdot} \epsilon {\cdot} \left[
         \fa \nna - \fb \nnb 
    \right]
    + \sqrt{8} \eta, \label{eomTheta}
\end{align}
where
$\rra = \RR + \frac{1}{2} \nn$, $\rrb = \RR - \frac{1}{2} \nn$,
$\epsilon_{yx} = -\epsilon_{xy} = 1$, $\epsilon_{xx} = \epsilon_{yy} = 0$,
$\xxi$ and $\eta$ are a random Gaussian vector and number, respectively, with $\left< \xxi(t) \right> = 0$, $\left< \xxi(t) \xxi(t') \right> =  \I \delta(t-t')$,
and $\left< \eta(t) \right> = 0$, $\left< \eta(t) \eta(t') \right> = \delta(t-t')$.
The free parameters in this study are the swim force $\fs(\rr)$ and the two angles $\phi_1$ and $\phi_2$.

The Fokker-Planck equation  corresponding to the SDEs
\ref{eomRd} and \ref{eomTheta} governs the time evolution of the probability density
$P(\RR, \theta, t)$ \citep{risken1996fokker}.
We coarse grain this equation by integrating out $\theta$ and only retain terms up to order $\sim \mathcal{O}\left( \nabla^2\right)$.
This results in a Fokker-Planck equation for the probability density of the dimer
$\rho(\RR,t) = \frac{1}{2 \pi} \int d \theta P(\RR, \theta, t)$.
In the following we only consider steady-state properties.
From the Fokker-Planck equation one can extract the steady-state density $\rho(\RR)$, flux $\JJ(\RR)$ and polarization $\pp(\RR) = \rho^{-1}(\RR) \int d \theta \nn P(\RR)$.
Details of the coarse graining procedure and the calculation of the steady-state properties are shown in the \cmnt{Appendix}.

\section{Results and Discussion}

Before we discuss the solution to the FPE, we inspect Eqs. \ref{eomRd} and
\ref{eomTheta}
to understand what kind of behavior one can expect from this system.
To do this, we ignore terms $\sim \mathcal{O}(\nabla^2 \fs)$, and assume that the swim force only depends on the $x$ coordinate.
Equations \ref{eomRd} and \ref{eomTheta} then  become 
\begin{align}
\dt x =&
	2 \fs(x) \left( \ccp \nx -  \ssp \ny \right)
	\nonumber \\
	&~~ - \left(\ccm \nx - \ssm \ny \right) \nx \dx \fs(x)
	+ \sqrt{2} \xi_x
\label{dx_long}
\\
\dt y =&
	2 \fs(x) \left( \ssp \nx + \ccp \ny \right)
	\nonumber \\
	&~~- \left(\ssm \nx + \ccm \ny \right) \nx \dx \fs(x)
	+ \sqrt{2} \xi_y
\label{dy_long}
\\
\dt \theta =&
	-4 \ssm \fs(x) - 2 \ssp \nx \dx \fs(x) + \sqrt{8} \eta,
\label{dtheta_long}
\end{align}
where $x = \RR \cdot \hat{\mathbf{e}}_x$,
$y = \RR \cdot \hat{\mathbf{e}}_y$, 
$\ccpm  = \cos(\phi_1) \pm \cos(\phi_2)$, and
$\sspm  = \sin(\phi_1) \pm \sin(\phi_2)$.

\cmnt{
Because of the torque on the orientation of the dimer,
that is the $-4\ssm \fs$ term in the Eq. \eqref{dtheta_long},
these dimers are chiral active particles \cite{lowen2016chirality,bechinger2016active},
and because of that they are odd diffusive \cite{hargus2021odd},
meaning that they have diffusive fluxes perperdicular to density gradients
(see App. \ref{app:odd_diffusion}).
}

\begin{figure}[t]
\centering
\includegraphics[width=0.8\columnwidth]{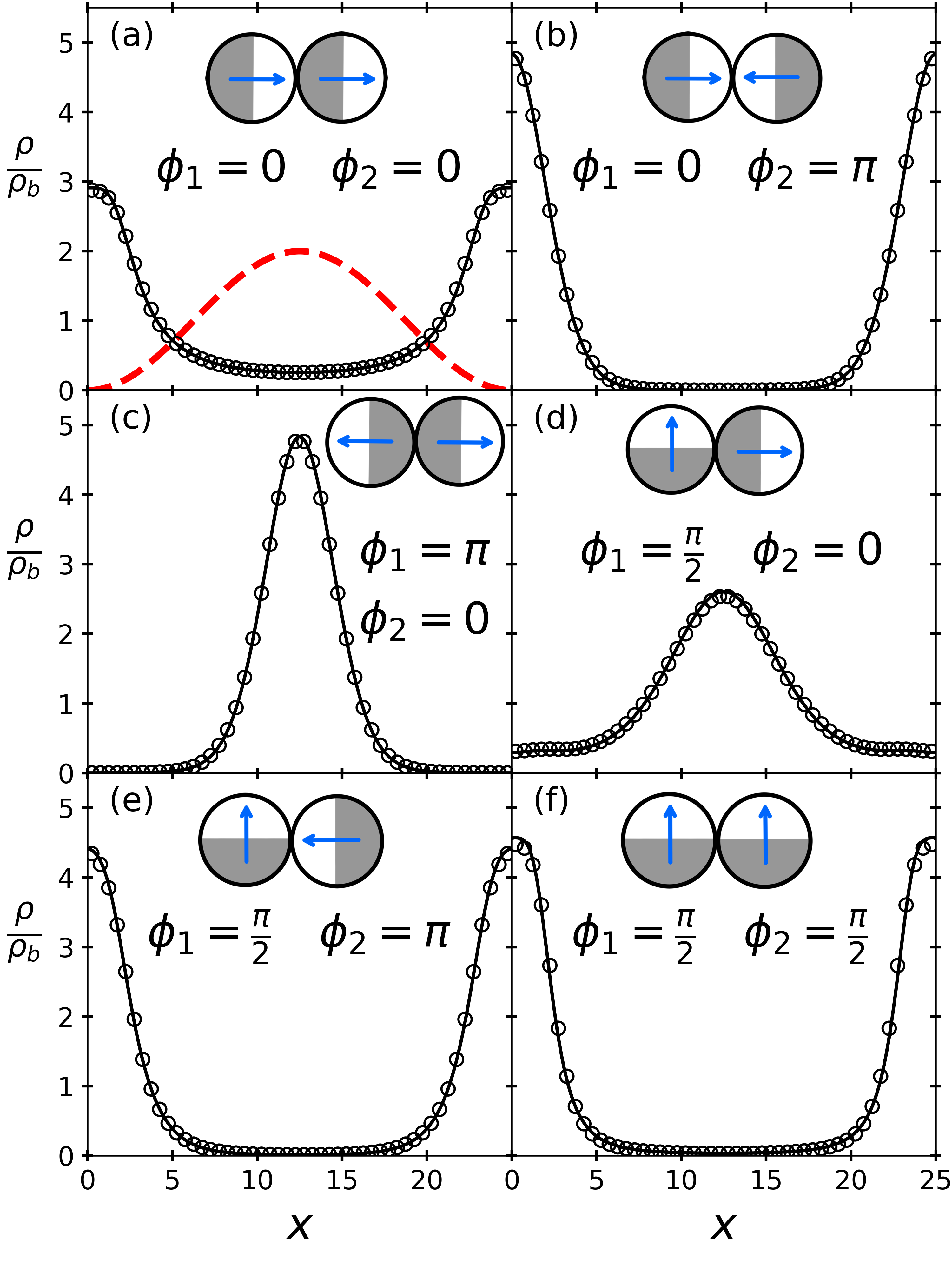}
\caption{Density for different dimers (see insets) relative to the bulk density 
$\rho_b = \int_0^Ldx \rho(x)/L$, where $L=25$ is the simulation box with periodic boundary conditions.
The orientations of the particles in the dimer are indicated in the figure.
The symbols show the simulation of Eqs. \eqref{eomRd} and \eqref{eomTheta},
the solid line show the theoretical prediction (Eq. \eqref{eq:U}),
and the red dashed line in (a) shows the shape of the swim-force profile 
$\fs(x) = 8  \left[ 1 + \sin\left(2 \pi x /L + 3 \pi/2\right) \right]$.
The orientation of the particles in the dimer can be used to control wether the 
dimer accumulates in regions where $\fs$ is small (panels a,b, e and f),
or in regions where $\fs$ is large (panels c and d).
}
\label{fig:density_structure}
\end{figure}

In order to get a better physical understanding of the different contributions to the equations of motion, a few illustrative examples are discussed.
A dimer with $\phi_1= \phi_2 = 0$,
shown in the inset of Fig. \ref{fig:density_structure} (a),
is structurally similar to a single ABP. Accordingly, this dimer accumulates where the swim force is small.
Dimers where the two active particles have opposite orientations along the orientation vector are shown in the insets of Fig. \ref{fig:density_structure} (b) and (c).
These dimers are symmetric under $\nx \rightarrow - \nx$.
Since the swim-force varies only along the $x$ coordinate, at any location,
a dimer with $\phi_1=0, ~ \phi_2 = \pi$,
experiences a net force towards the region of small swim force (antichemotactic) whereas an dimer with $\phi_1=\pi, ~ \phi_2 = 0$ experiences a net force towards the region of large swim force (chemotactic). 

A particularly interesting structure is an dimer with $\phi_1 = \phi_2 = \pi /2$ shown in the inset of Fig. \ref{fig:density_structure} (f).
In this case, the orientations of the two particles are parallel to each other and perpendicular to the orientation vector.
The equations of motion are
\begin{align}
\dt x =&
  -4 \ny  \fs(x) + \sqrt{2} \xi_x,
\\
\dt y =&
	4  \nx \fs(x) + \sqrt{2} \xi_y,
\\
\dt \theta =& - 4 \nx \dx \fs(x) +  \sqrt{8} \eta.
\end{align}
The first two of these equation are the same as that for a ABP with rotated orientation vector.
In the equation of motion of the angle, however, a new feature appears.
There is an active torque on the dimer prortional to $\nx$ and the gradient in the 
swim force.
This torque rotates the dimer, like a steering wheel, in such a way that the orientation vector points in the direction perpendicular to the gradient in the swim force,
therefore, this torque stabilizes the dimer such that the active forces point in the direction opposite the gradient in the swim force.
Accordingly, this dimer accumulates where the swim force is small.

\begin{figure}[t]
\centering
\includegraphics[width=0.9\columnwidth]{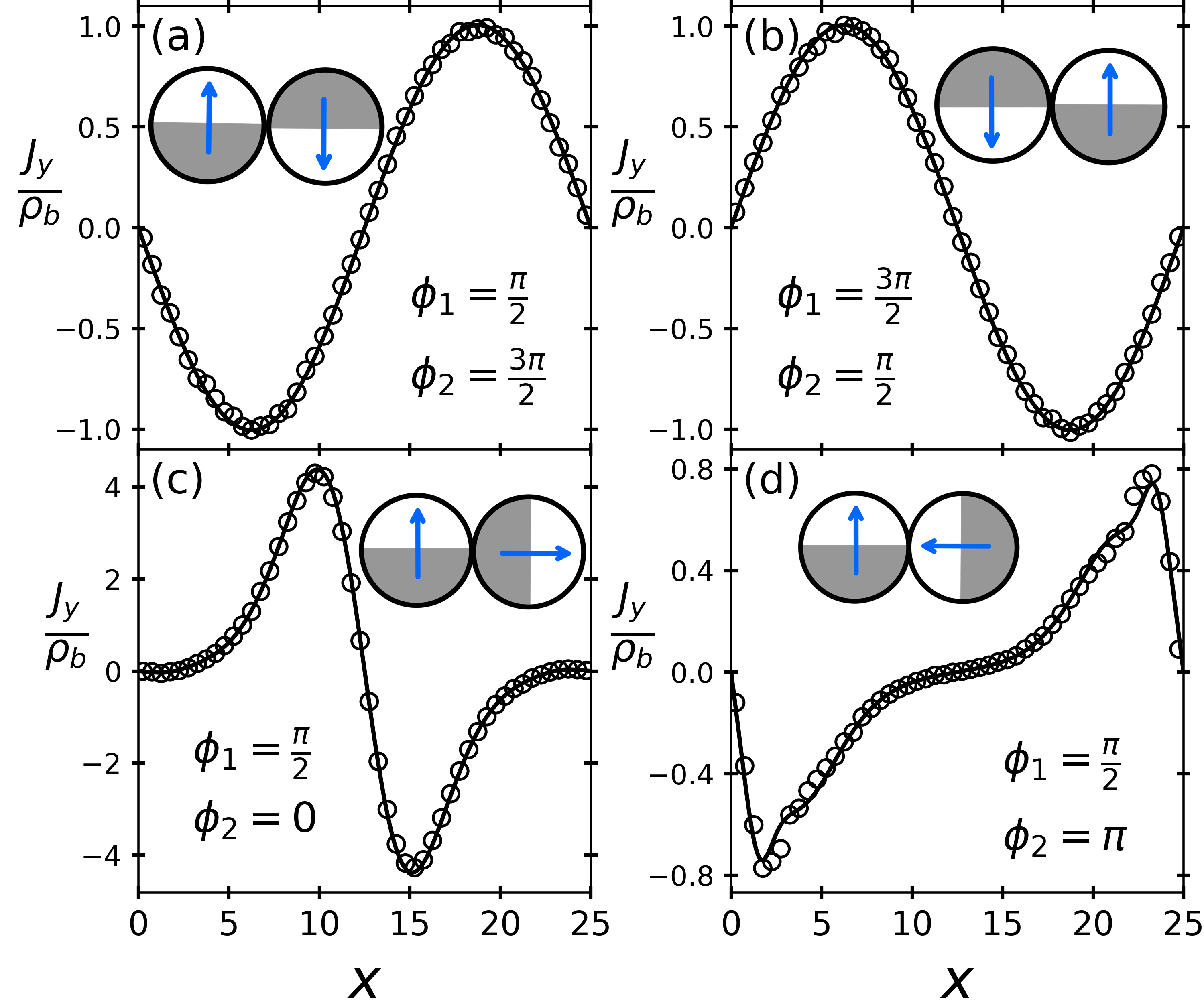}
\caption{
Flux perpendicular to swim-force gradients. The bulk density is $\rho_b = \int_0^L dx \rho(x)/L$ with $L=25$.
\cmnt{
These transverse fluxes are reminiscent of, for instance,
chemotactic sea urchin sperm swimming in the presence of a chemical source \cite{friedrich2007chemotaxis}.}
The swim force is
$\fs(x) = 8 \left[ 1 + \sin\left(2 \pi x /L + 3 \pi/2\right) \right]$
(same as Fig \ref{fig:density_structure}).
The orientations of the particles in the dimer are shown in the figures.
}
\label{fig:odd_diffusive}
\end{figure}

Dimers in which the orientations of the two active particles have opposite orientations and perpendicular to the orientation vector are shown in the insets of Fig. \ref{fig:odd_diffusive} (a) and (b). For $\phi_1 = \pi /2,  \phi_2 = 3\pi/2$, the equations of motion are
\begin{align}
\dt x =&
   2 \ny \nx \dx \fs(x) + \sqrt{2} \xi_x,
\\
\dt y =&
	-2 \nx^2 \dx \fs(x) + \sqrt{2} \xi_y,
\\
\dt \theta =& -8 \fs + \sqrt{8} \eta.
\end{align}
Two features of these equations are noteworthy.
Firstly, there is an active torque acting on the dimer (the $-8 \fs$ part in the equation for the time evolution of the angle).
This is equivalent to the active torque in case of an active chiral particle
\citep{kummel2013circular,lowen2016chirality,caprini2019active}.
Secondly, the term $-\nx^2 \dx \fs(x)$ in the time evolution equation for the $y$ coordinate is nonzero on average for a fixed value of $x$. 
Since there is translational invariance in the $y$ direction, this effective force, remains unbalanced giving rise to stationary fluxes perpendicular to the swim-force and density gradients.
Note that since the dimer is symmetric under $\nx \rightarrow - \nx$, on average the $x$ coordinate gets no contribution from the swim-force gradients. Accordingly, this dimer shows no preferential accumulation in a swim-force gradient. The behaviour of a dimer with $\phi_1 = 3\pi /2,  \phi_2 = \pi/2$ (inset of Fig. \ref{fig:odd_diffusive} (b)) is similar except that its chirality is reversed.

The structural properties of the dimer, namely net activity proportional to $\fs$, force proportional to $\nabla \fs$, torque proportional to $\fs$, and a torque proportional to $\nabla \fs$,
are determined by the orientation of the two particles
and result in two classes of steady-state behaviour.
One could design the dimer in such a way that it preferentially moves towards regions with high or low swim force.
Going beyond the examples above, for a generic structure of the dimer,
the stationary density distribution can be obtained from the coarse grained Fokker Planck equation by setting the flux along the gradient of swim force to zero .
\cmnt{
(see the Appendix for the derivations of Eqs. \eqref{eomRd}
\eqref{eomTheta} and \eqref{eq:U}).
}
This yields
\begin{align}
 \rho(x) \propto \exp(-U),
\end{align}
 with
\begin{align}
 U = \frac{c}{2 d} \fs
    + \frac{b}{4 d} \ln\left( 1 + d \fs^2 \right)
    + \frac{a d - c}{2 d^{3/2}} \atan\left( \sqrt{d} \fs \right),
    \label{eq:U}
\end{align}
 where $a=\ccm$, $b= \ccp^2 + 2\ssp^2$, $c=\ssm^2 \ccm - \ssm \ssp \ccp$,
 and $d= \frac{1}{2}\left( 2 \ssm^2 + \ccp^2 + \ssp^2 \right)$. Figure~\ref{fig:density_structure} shows the stationary density distribution of dimers with different structures obtained from simulations of Eqs.~ \eqref{eomRd} and \eqref{eomTheta}. Depending on the structure, dimers accumulate in the regions where swim force is small or large.
The theoretical predictions (Eq.~\eqref{eq:U}) are in perfect agreement with the simulations. 
 
The steady-state density distribution obtained from the coarse grained Fokker Planck equation is Boltzmann-like with an effective potential ($U$). However, this does not imply that on this coarse-grained level the dynamics obey detailed balance; there are configurations of the dimer that result in steady-state fluxes in the direction perpendicular to gradients in the swim force (see Fig. \ref{fig:odd_diffusive}). For instance, a dimer with $\phi_1 = \pi/2$ and $\phi_2 = 3\pi/2$  is a chiral particle that rotates clockwise whereas a dimer with $\phi_1 = 3\pi/2$ and $\phi_2 = \pi/2$ rotates anticlockwise.
While these dimers show no preferential accumulation ($U=0$ in Eq.~\eqref{eq:U}), the gradient in the swim force gives rise to a net force along the $y$ direction that gives rise to fluxes $J_y = -\rho_b\partial_x f_s$  for the clockwise dimer (Fig.~\ref{fig:odd_diffusive}(a)) and $J_y = \rho_b\partial_x f_s$ for the anticlockwise dimer (Fig.~\ref{fig:odd_diffusive}(b)).
In case of a generic dimer structure, for which the stationary distribution is not homogeneous (Fig.~\ref{fig:odd_diffusive}(c-d)), the flux along the $y$ direction is perpendicular to the density gradient (along $x$).
Fluxes perpendicular to the density gradients is a characteristic property of odd-diffusive systems~\cite{vuijk2020lorentz,abdoli2020correlations,hargus2021odd}.
The odd-diffusive flux of active dimers can be obtained from the coarse grained Fokker Planck equation \cmnt{(see the Appendix)}. These fluxes are a clear indication of broken detailed balance, and show that on this coarse-grained level not all properties of the dimer can be captured by the effective potential (Eq.~\ref{eq:U})
alone.

\begin{figure}[t]
\includegraphics[width=\columnwidth]{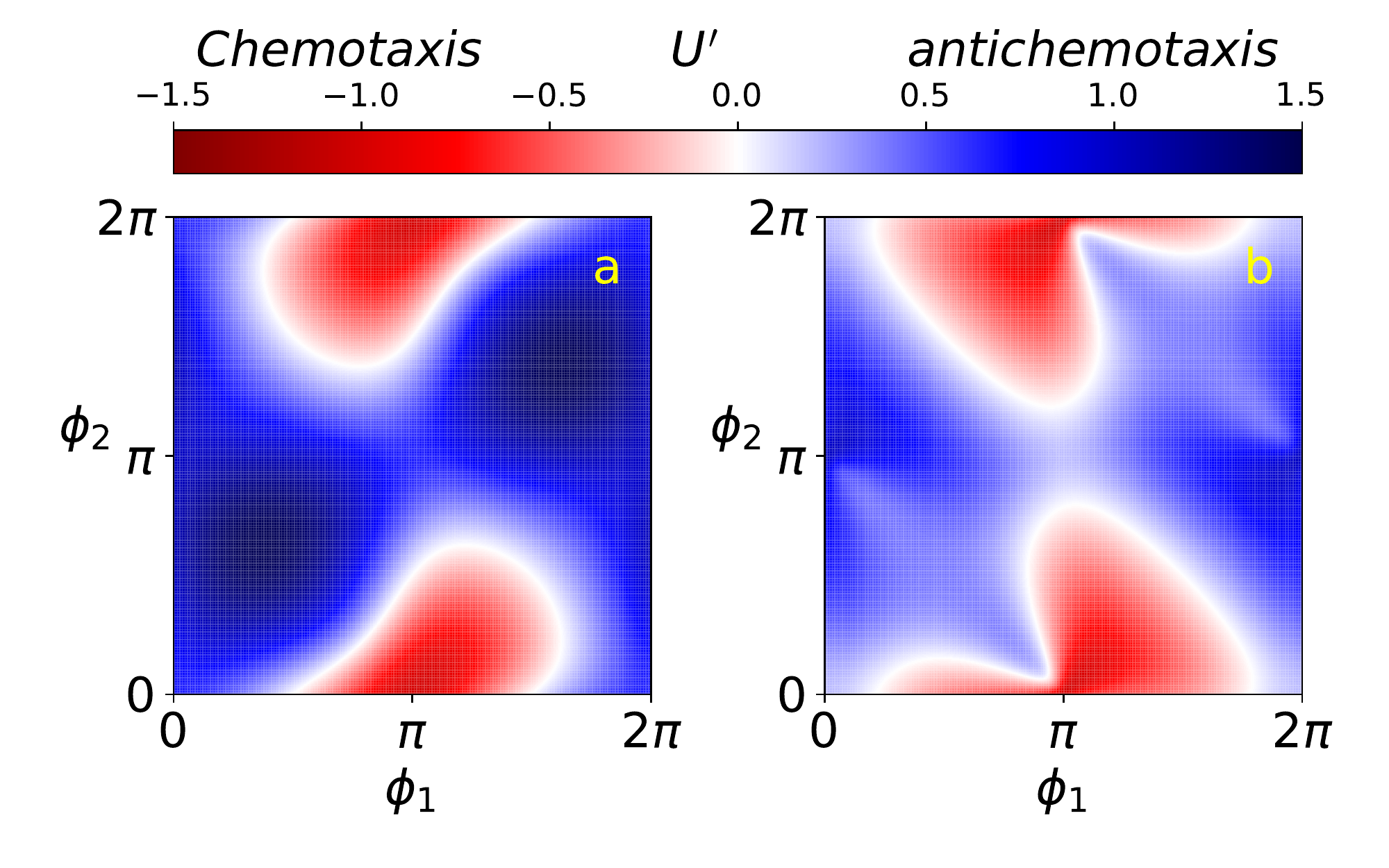}
\caption{
Phase diagrams for the tactic behaviour of dimers for a value of the swim force of $f_s = 0.5$ in (a) and $f_s = 5$ in (b).
Every point in the $\phi_1$-$\phi_2$ plane corresponds to a different dimer structure. The blue region corresponds to antichemotactic dimers,
which experience an effective force down the swim-force gradients ($U' > 0$).
The red region corresponds to chemotactic dimers ($U' <0$).
These tactic regions are separated by white boundaries which correspond to dimers which show no preferential accumulation ($U' = 0$).
The phase behaviour is dependent on the magnitude of the swim force implying that the same dimer can be chemotactic or antochemotactic depending on the magnitude of the swim force. }
\label{fig:phase_diagram}
\end{figure}

\begin{figure}[t]
\centering
\includegraphics[width=0.7\columnwidth]{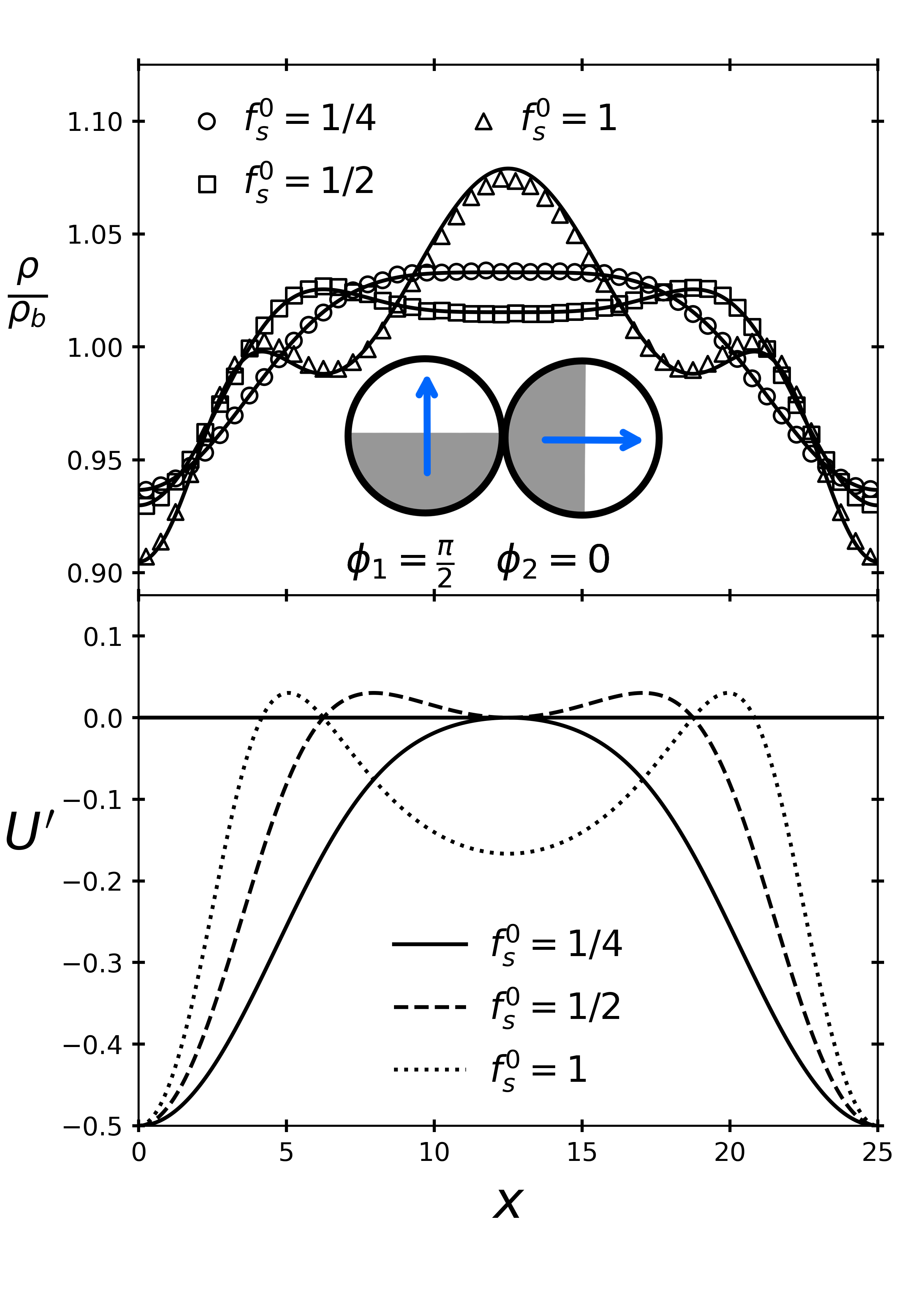}
\caption{ Top: Density relative to the bulk density for a dimer with $\phi_1=\pi/2$ and $\phi_2=0$ (see inset) for different values of the swim force.
This is the same dimer as Fig. \ref{fig:density_structure} d.
The bulk density is $\rho_b = \int_0^L dx \rho(x)/L$,
where $L=25$ is the simulation box with periodic boundary conditions.
The swim force is
$\fs(x) = \fs^0 \left[ 1 + \sin\left(2 \pi x /L + 3 \pi/2\right) \right]$,
with the value of $\fs^0$ as indicated in the legend.
Bottom: Derivative of the effective potential $U' = d U/ df_s$.
Wherever $U'<0$ ($U'>0$) the dimer is chemotactic (antichemotactic).
The density profile changes qualitatively when the swim force increases.
In this case there is a single peak in the density where $\fs$ is large for 
$\fs^0=1/4$,
as the swim force increases ($\fs^0=1/2$) the peak split in two,
and if the swim force is increased further ($\fs^0=1$) a third peak appears.
}
\label{fig:density_magnitude}
\end{figure}
Whether a dimer accumulates in small or large swim-force regions is determined by the effective force that it experiences in swim-force gradients. The effective force can be obtained from the effective potential as $ - \nabla U =  - U' \nabla f_s$, where $U' = \frac{dU}{df_s}$.
Wherever $U'< 0$ ($U'>0$) the dimer moves up (down) the swim-force gradient,
corresponding to chemotactic (antichemotactic) behaviour.
Figure~\ref{fig:phase_diagram}(a) shows a phase diagram in the $\phi_1$-$\phi_2$ plane that categorises different dimer structures according to their tactic behaviour for $f_s = 0.5$.
Dimers which show no preferential accumulation in a swim-force gradient correspond to the white lines shown in Fig.~\ref{fig:phase_diagram}(a) obtained as $U'=0$.

As can be seen in Fig.~\ref{fig:phase_diagram}(b),
the phase behaviour depends on the magnitude of the swim force.
This means that a dimer can be antichemotactic in case of a small swim force and chemotactic in case of a large swim force, or vice versa.
This can result in "local" (anti)chemotaxis, as shown in Fig.~\ref{fig:density_magnitude},
where at low swim force there is a single peak in the density that coincides with the peak in $\fs$,
as the swim force increases, the peak splits in two,
and on further increasing the swim force a third peak appears.
The density distribution has multiple peaks because $U'$ is negative in some regions and positive in others which can be regarded as coexisting chemotactic and antichemotactic dynamic phases.

Using experimental values from Ref. \cite{ebbens2010selfassembled}
for dimers with a constant activity,
we estimate the change in density relative to a passive region for the dimer shown in Fig. \ref{fig:density_structure} (d)
to be $\rho_{active}/\rho_{passive} \approx 4$
\cmnt{(see App. \ref{app:mapping_experimental_dimer} for details of the estimate)}.
While this is a conservative estimate, it is likely possible to obtain much larger density changes,
for example for the dimer in Fig. \ref{fig:density_structure} (b),
or by different experimental conditions.
To obtain better predictions for experimental system, it would be interesting 
to include the effects of the specific self-propulsion mechanism \cite{sanchez2015chemically} and the hydrodynamic interaction between the two particles in the dimer \cite{reigh2015catalytic,reigh2018diffusiophoretically}.

\begin{acknowledgments}
A.S. acknowledges support by the Deutsche Forschungsgemeinschaft (DFG) within the project SH 1275/3-1.
We thank H. L{\"o}wen for fruitful discussions.
\end{acknowledgments}

\appendix

\section{Active dimer model}\label{app:model}
The equations of motion for the dimer (see Fig. 1 in the main text) are
\begin{align}
\dt \rra &=
  ~~  \frac{1}{\gamma} \FF + \frac{1}{\gamma} \fa \nna + \sqrt{2T/\gamma} \xxia,
    \label{app:eomr2}
\\
\dt \rrb &=
 -   \frac{1}{\gamma} \FF + \frac{1}{\gamma} \fb \nnb + \sqrt{2T/\gamma} \xxib,
    \label{app:eomr1}
\end{align}
where 
$\rr_1$, $\nna$, $\rr_2$, and $\nnb$ are, respectively, the
position and orientation vectors of particles 1 and 2,
$\gamma$ is the friction constant of a single particle in the dimer,
$\fs(\rr)$ is the swim force, 
$T$ is the temperature in units such that the Boltzmann constant is unity,
the vectors $\xxia$ and $\xxib$ are random Gaussian vectors with
$\left<\xxia(t)\right> = \left< \xxib(t) \right> = 0$ and
$\left<\xxia(t) \xxia(t') \right> = \left<\xxib(t) \xxib(t') \right> = \I \delta(t-t')$.
The force between the two active particles $\FF$ is always such that the distance between the two is constant and equal to $l$.

We make the equations dimensionless by $\rra \rightarrow  l \rra$, $\rrb \rightarrow l \rrb$,
$t \rightarrow 2 \gamma l^2 t /  T$ and  $\fs(\rr) \rightarrow 2 T  \fs(\rr)/l$, so length is measured in units such that the distance between the two particles is unity, time is measured in units such that the dimer typically diffuses a unit length per unit time, and forces are measured in units of $2T/l$.
The dimensionless equations corresponding to Eqs. \eqref{app:eomr2} and \eqref{app:eomr1} are
\begin{align}
\dt \rra &=
  ~~  4 \FF + 4 \fa \nna + 2 \xxia, \label{app:eomr1d}
\\
\dt \rrb &=
 -   4 \FF + 4 \fb \nnb + 2 \xxib. \label{app:eomr2d}
\end{align}

These equation can be rewritten using the center-of-mass coordinate $\RR = \frac{1}{2} \left( \rra + \rrb \right)$
and the relative coordinate $\rr = \rra - \rrb$:
\begin{align}
\dt \RR &=
    2 \left[ \fa \nna + \fb \nnb \right] + \sqrt{2} \xxi,
\label{app:eomRd}
\\
\dt \rr &=
  -8 \FF -   4 \left[ \fa \nna - \fb \nnb \right] + \sqrt{8} \eeta,
\end{align}
where $\left<\xxi(t)\right> = \left<\eeta(t) \right> = 0$ and
$\left<\xxi(t)\xxi(t')\right> = \left<\eeta(t) \eeta(t') \right> = \I \delta(t-t')$.
The second equation accounts for the relative movement of the two particles,
which can be decomposed in a rotation of the unit vector pointing from
$\rra$ to $\rrb$, and a change in the distance between the two particles
(see Sec. \ref{torque_decomposition}).
$\rr = r \nn$, with $\nn = \left( \cos(\theta), \sin(\theta) \right)$.
The Langevin equation for the change in the distance is
\begin{align}
\dt r = -8 \nn {\cdot} \FF -
    4 \nn {\cdot} \left[ \fa \nna - \fb \nnb \right]
    + \sqrt{8} \nn {\cdot} \eeta.
\end{align}
Because the particles are connected by a stiff rod, the force due to the rod $\FF$ is always such that $r=1$ and $\dt r = 0$.
The equation for the orientation is
\begin{align}
\dt \nn =&
-8 \left( \I - \nn \nn \right) {\cdot} \FF
 - 4 \left( \I - \nn \nn \right) {\cdot} \left[
        \fa \nna - \fb \nnb 
    \right]
\nonumber \\
&+ \sqrt{8} \left( \I - \nn \nn \right) {\cdot} \eeta.
\end{align}
This equation should be integrated with the Stratonovich rule.
The first term on the right-hand side is zero because $\FF\propto \nn$, and the last term can be replaced by $\epsilon {\cdot} \nn \eta$ where 
$\epsilon_{yx} = - \epsilon_{xy} = 1$ and $\epsilon_{xx} = \epsilon_{yy} = 0$,
so
\begin{align}
\dt \nn =
  4 \left( \I - \nn \nn \right) {\cdot} \left[
        \fa \nna - \fb \nnb 
    \right]
+ \sqrt{8} \epsilon {\cdot} \eeta.
\label{app:eomn}
\end{align}

The previous equation is equivalent to
\begin{align}
\dt \theta
&=
    4 \nn {\cdot} \epsilon {\cdot} \left[
         \fa \nna - \fb \nnb 
    \right]
    + \sqrt{8} \eta, 
\label{app:eomTheta}
\end{align}
with $\left< \eta(t) \right> = 0$ and $\left< \eta(t) \eta(t') \right> = \delta(t-t')$.
This equation together with Eq. \eqref{app:eomRd} describes the dynamics of the dimer and are used for the simulations.

\section{Small gradient approximation}\label{app:small_gradient}

The orientation vectors of the active particles can be written as a rotation of the orientation vector of the dimer: $\nna= R_1 \nn$ and $\nnb= R_2 \nn$,
where $R_1 = R(\phi_1)$, $R_2 = R(\phi_2)$ and
\begin{align}
R(\phi) =
\begin{bmatrix} 
    \cos \phi & - \sin{\phi} \\
    \sin \phi &  \cos{\phi}
\end{bmatrix}.
\end{align}
We define
$A = R_1 + R_2 = \ccp \I + \ssp \epsilon$
and $B = R_1 - R_2 = \ccm \I + \ssm \epsilon$,
where $\ccpm = \cos(\phi_1) \pm \cos(\phi_2)$ and
$\sspm = \sin(\phi_1) \pm \sin(\phi_2)$.

We assume gradients in the swimforce to be small, so we expand the swim force in Eqs. \eqref{app:eomRd} and \eqref{app:eomTheta}:
\begin{align}
\dt \RR &=
 2 \fs A {{\cdot}} \nn
-  \nn {{\cdot}} (\nabR \fs) B {\cdot} \nn
+ \sqrt{2} \xxi
+ \order(\nabR^2 \fs), 
\label{app:eomRd1}
\\
\dt \theta &=
      4 \nn {\cdot} \epsilon {\cdot} B {\cdot} \nn \fs
    - 2 \nn {\cdot} \epsilon {\cdot} A {\cdot} \nn ~ \nn {\cdot} \nabR \fs
    + \sqrt{8} \eta
	\nonumber \\
	&~~~~
    + \order(\nabR^2 \fs),
\nonumber
\\
 &=
    - 4 \ssm \fs
    + 2 \ssp \nn {\cdot} \nabR \fs
    + \sqrt{8} \eta
    + \order(\nabR^2 \fs), 
\label{app:eomTheta1}
\end{align}
where $\fs= \fs(\RR)$, and $\nabR$ is the gradient with respect to $\RR$.

\cmnt{
The motion of a single chiral active Brownian particle is described by
\cite{lowen2016chirality,bechinger2016active}
\begin{align}
\dt \rr &= v_s \nn + \sqrt{2D} \xxi,\\
\dt \theta &= \Omega + \sqrt{2 D_r} \eta,
\end{align}
where $v_s$ is the swim speed, $D$ the thermal diffusion constant,
$\Omega$ the toque on the particle and $D_r$ its rotational diffusion constant.
Comparing with Eqs. \eqref{app:eomRd1} and \eqref{app:eomTheta1} shows that,
if the swim force is constant,
the swim speed of the dimer is
$v_s = | 2 \fs A\cdot\nn| = 2 \fs \sqrt{\ccp^2 + \ssp^2}$,
its thermal diffusion constant is $D = 1$,
the torque is $\Omega = -4 \ssm \fs$ and the rotational diffusion constant is
$D_r = 4$.
The active diffusion constant is 
\begin{align}
D_a = \frac{v_s^2}{2D_r}
= \frac{1}{2} \fs^2 \left( \ccp^2 + \ssp^2 \right)
\end{align}
If $\Omega \neq 0$ there is a torque on the particle, and it swims in circles.
These particles are called \emph{circle swimmers} \cite{kummel2013circular} or \emph{chiral active particles} \cite{caprini2019active}.
The chirality results in diffusive fluxes perperndicular to density gradients.
This property is called odd diffusion
(see App.\ref{app:odd_diffusion}).
By tuning $\ssm$ one can tune the chirality and with that the odd diffusivity of these dimers.
}

\cmnt{
The Fokker-Planck equation corresponding to  Eqs. \eqref{app:eomRd1} and \eqref{app:eomTheta1} is}
\begin{align}
\dt P = & 
    - 2 \nabR {\cdot} \left[ \fs A {\cdot} \nn P \right]
    +  \nabR {\cdot} \left[ B {\cdot} \nn ~ \nn {\cdot} (\nabR\fs) P \right]
	\nonumber \\ &
    + \nabR^2 P
    + 4 \ssm \fs \rot P - 2 \ssp \rot \left[ \nn {\cdot} (\nabR \fs) P \right]
	\nonumber \\ &
    + 4 \rot^2 P
    + \mathcal{O}\left(\nabR^3\right),
\\
=&
   -2 A_{ij} \nabR_i \left[ \fs \n_j P \right]
    + B_{ij} \nabR_i \left[ (\nabR {}_k \fs) \n_j \n_k P \right]
	\nonumber \\ &
    + \nabR^2 P
    +4 \ssm \fs \rot P 
    -2 \ssp (\nabR {}_i \fs) \rot \left[ \n_i P \right]
	\nonumber \\ &
	 + 4 \rot^2 P
    + \mathcal{O}\left(\nabR^3\right),
\label{app:FPE}
\end{align}
where
$\rot = \frac{\partial}{\partial \theta} = n_x \frac{\partial}{\partial n_y} - n_y\frac{\partial}{\partial n_x}$.
The previous equation is valid up to third order in the gradient operator because the SDEs (Eq. \eqref{app:eomRd1} and \eqref{app:eomTheta1}) are only valid up to second order in the gradient.

We expand the probability density in eignefunctions of $\rot^2$ \citep{cates2013when,duzgun2018active,saintillan2015theory}:
\begin{align}
P(\RR, \nn(\theta), t) =&
\rho(\RR,t)
+ \ssig(\RR, t) {\cdot} \nn
\nonumber \\ &
+ \tau(\RR, t) {:} \left( \nn \nn - \I /2 \right)
+ \Theta,
\end{align}
where $\rho(\RR,t = \frac{1}{2 \pi}\int d \theta P(\RR,t)$ is the density,
$\ssig$ and $\tau$ are the coefficients of, respectively, the second and third moment,
and $\Theta$ is the projection onto higher-order moments.
If the swim force is uniform ($\nabR \fs = 0$),
the system is isotropic and therefore $\ssig=0$, $\tau=0$ and $\Theta =0$,
all moments except the zeroth are at least proportional to $\nabla$.
For the integral over the orientation $\nn(\theta)$ we write
$\left< \boldsymbol{\cdot} \right> = \frac{1}{2 \pi} \int_0^{2\pi} d \theta ~ \boldsymbol{\cdot}$.
The equation for $\rho$ can be obtained by averaging Eq. \eqref{app:FPE}:
\begin{align}
\dt \left< P(t) \right>
=&
    - 2 A_{ij} \nabR_i \left[ \fs \left< n_j P \right> \right]
	\nonumber \\ &
    + B_{ij} \nabR_i \left[ \left( \nabR_k \fs \right) \left< n_j n_k P \right> \right]
    + \nabR^2 \left<P\right> \nonumber \\
   & 
    + 4 \ssm \fs \left< \rot P \right>
    - 2 \ssp \left( \nabR_i \fs \right) \left< \rot P n_i \right>
	\nonumber \\ &
    + 4 \left< \rot^2 P \right>
    + \mathcal{O}\left(\nabR^3\right).
\end{align}
All averages with $\rot$ in this equation are zero.
With 
$\left<P\right> = \rho$,
$\left< n_j P\right> = \sigma_j /2$ and
$\left< n_j n_k P \right> = \delta_{jk} \left<P\right>
+\left< \left( n_j n_k - \delta_{jk}/2 \right) P \right>
= \delta_{jk} \rho /2 + \tau_{jk}/4 $
the previous equation becomes
\begin{align}
\dt \rho 
=& - \nabR {\cdot} \JJ,
\label{app:rho}
\end{align}
with
\begin{align}
J_i =&
    A_{ij} \fs \sigma_j
    - \frac{1}{2} B_{ij} \left( \nabR_j \fs \right) \rho 
    - \frac{1}{4} B_{ij} \left( \nabR_k \fs \right) \tau_{jk}
	\nonumber \\ &~~
    - \nabR {}_i \rho
    +\mathcal{O} \left( \nabR^2 \right),
    \nonumber \\
=&
    A_{ij} \fs \sigma_j
    - \frac{1}{2} B_{ij} \left( \nabR_j \fs \right) \rho 
    - \nabR_i \rho 
    +\mathcal{O} \left( \nabR^2 \right),
\label{app:J}
\end{align}
where in the last step we ignored the term with $\tau$ because if there is no gradient in the swim force,
the system is isotropic, so there is no nematic ordering ($\tau=0$),
and therefore $\tau \sim \mathcal{O}\left( \nabR \fs \right)$
and $ \tau_{jk} \nabR {}_k \fs \sim \mathcal{O}\left( \nabR^2 \right)$.

To get  an equation for $\ssig$, we multiply Eq. \eqref{app:FPE} by $n_l$ and average over $\nn$:
\begin{align}
2 \dt \left< n_l P \right>
=&
    - 2 A_{ij} \nabla_i \left[ \fs 2 \left< n_j n_l P \right> \right]
	\nonumber \\ &
    + B_{ij} \nabla_i \left[ \left( \nabR_k \fs \right) 2 \left< n_j n_k n_l P \right>\right]
	\nonumber \\ &
    + \nabR^2 2 \left< n_l P\right>
	+ 4 \ssm \fs 2 \left< n_l \rot P \right>
	\nonumber \\ &
    - 2 \ssp \left(\nabla_i \fs \right) 2 \left< n_l \rot  n_i P \right>
	\nonumber \\ &
    + 8 \left< n_l \rot^2 P \right>
    + \mathcal{O}\left(\nabR^3\right).
\end{align}
With what we used before together with
$2 \left< n_j n_k n_l P \right> = 2 \sigma_m \left< n_j n_k n_l n_m  \right>
    + \left< n_j n_k n_l \Theta \right>
= \sigma_m T^{(4)}_{jklm} + \chi_{jkl}$
where $T^{(4)}_{jklm} = \left( \delta_{jk} \delta{lm} + \delta_{jl} \delta_{km}
    + \delta_{jm} \delta_{kl} \right)$
and $\chi_{jkl}$ is the projection of $n_j n_k n_l $ on $\Theta$,
$2\left<n_l \rot P\right> = -2\left< \left( \rot n_l \right) P \right>
    = - \epsilon_{lm} 2 \left< n_m P \right>
    = - \epsilon_{lm} \sigma_m$,
$2\left<n_l \rot n_i P \right> = -2 \left< \left( \rot n_l \right) n_i P \right>
    = - \epsilon_{lm} 2 \left< n_m n_i P\right>
    = - \epsilon_{lm} 2 \left( \frac{1}{2} \delta_{mi} \rho + \frac{1}{4} \tau_{mi} \right)
    = - \epsilon_{li} \rho - \frac{1}{4} \epsilon_{lm} \tau_{mi}$,
and $ 2 \left< n_l \rot^2 P \right> = 2 \left< \left( \rot^2 n_l \right) P \right>
    = - 2 \left< n_l P \right> = - \sigma_l$,
the previous equation becomes
\begin{align}
\dt \sigma_l =&
    - 2 A_{il} \nabR_i \left[ \fs \rho \right]
    - A_{ij} \nabR_i \left[ \fs \tau_{jl} \right]
	\nonumber \\ &
    + B_{ij} \nabR_i \left[ \left( \nabR_k \fs \right)
                \left( \sigma_m T^{(4)}_{jklm} + \chi_{jkl} \right) \right]
	\nonumber \\ &
    + \nabR^2 \sigma_l
	- 4 \ssm \fs \epsilon_{lm} \sigma_m
    + 2 \ssp \epsilon_{li} \left( \nabR_i \fs \right) \rho
	\nonumber \\ &
    +  \ssp \left( \nabR_i \fs \right) \epsilon_{lm}\tau_{mi}
    - 4 \sigma_l
    + \mathcal{O}\left(\nabR^3\right).
\end{align}
Equation \eqref{app:rho} shows that the time scale of the time evolution in the density is
$\sim \mathcal{O}\left( \nabR^{-1} \right)$. 
The previous equation show that the time scale of the time evolution of $\ssig$ is
at least $\sim 4$, so compared to $\rho$, $\ssig$ is a fast degree of freedom, 
and therefore $\dt \ssig \approx 0$.
With this together with
$\ssig \sim\mathcal{O}\left(\nabR\right)$,
$\tau\sim\mathcal{O}\left(\nabR\right)$,
$\chi \sim\mathcal{O}\left(\nabR\right)$,
we can re-arrange the previous equation:
\begin{align}
 \sigma_j =&
    - \frac{1}{2}
    \frac{1}{1 + \ssm^2 \fs^2}
    \left[
        \delta_{jl} - \ssm \fs \epsilon_{jl}
    \right]
	\nonumber \\ &
    \times \left[
        A_{kl} \nabR_k \left( \fs \rho \right)
        - \ssp \epsilon_{lk} \left( \nabR_k \fs \right) \rho 
    \right]
	\nonumber \\ &
    + \mathcal{O}\left( \nabRd^2 \right).
\label{app:sigma}
\end{align}
Equations \eqref{app:rho}, \eqref{app:J} and \eqref{app:sigma} describe the coarse-grained dynamics of the dimer.
We assume gradients in the swimforce to be small, so we expand the swim force in Eqs. \eqref{app:eomRd} and \eqref{app:eomTheta}:

\cmnt{
\section{Odd diffusion}\label{app:odd_diffusion}
Odd-diffusive systems have a diffusion tensor with antisymmetric components
which can be written as\cite{hargus2021odd,vuijk2020lorentz,kalz2022collisions,abdoli2020correlations}
\begin{align}
D_{ij} = D_{\parallel} \delta_{ij}
  + D_{\perp} \epsilon_{ij}.
\end{align}
The diagonal components of this tensor ($D_\perp$) are related to the diffusion along the density gradient;
the antisymmetric components ($D_\parallel$) are related to the diffusion perpendicular to the density gradient.

The diffusion tensor of the dimers can be calculated from Eqs. \eqref{app:J} and \eqref{app:sigma} resulting in
\begin{align}
D_{\parallel} &=
  1 + D_a\frac{1}{1 + \omega^2},\\
D_{\perp} &= D_a \frac{\omega}{1 + \omega^2},
\end{align}
where $\omega = \Omega/D_r = - \ssm \fs$ is the active torque divided by the rotational diffusion constant.
}
\section{Steady-State }\label{app:steady-state}
\subsection{Density}
We calculate the steady-state density for $\fs = \fs(x)$,
so $\rho=\rho(x)$ and $\ssig = \ssig(x)$.
In this case the flux in the $x$-direction is zero:
\begin{align}
0
=&
J_x \nonumber \\
=&
    \fs A_{xj} \sigma_j
    + \frac{1}{2} B_{xx} \left( \nabR {}_x \fs  \right) \rho
    - \nabR {}_x \rho.
\end{align}
For the first term we use Eq. \eqref{app:sigma}:
\begin{align}
A_{xj} \sigma_j
=&
    -\frac{1}{2} \frac{1}{1 + \ssm^2 \fs^2}
    \left(
        \ccp^2 + 2\ssp^2 - \ssm \ssp \ccp \fs 
    \right)
	\nonumber \\ &
	~~~~~\times
    \left( \nabR_x \fs \right) \rho
	\nonumber \\ &
     -\frac{1}{2} \frac{\fs}{1 + \ssm^2 \fs^2}
    \left( \ccp^2 + \ssp^2 \right)
    \nabR_x  \rho.
\end{align}
With this, the steady-state density becomes
\begin{align}
\rho \propto e^{-U},
\end{align}
where
\begin{align}
\nabR_x U & =
     \frac{1}{2} \left( \nabR_x \fs \right)
	\nonumber \\ &
    \times \frac{
        \ccm + \left(\ccp^2 + 2 \ssp^2\right) \fs 
        + \ssm\left( \ssm \ccm - \ssp \ccp \right) \fs^2
    }
    {
       1 +\frac{1}{2} \left( \ccp^2 + \ssp^2 + 2 \ssm^2 \right) \fs^2 
    },
\\
U = &
    \int d x ~  \nabR_x U ,
\nonumber  \\
=&
     \frac{c}{2 d} \fs
    + \frac{b}{4 d} \ln\left( 1 + d \fs^2 \right)
    + \frac{a d - c}{2 d^{3/2}} \atan\left( \sqrt{d} \fs \right),
\end{align}
where $a=\ccm$, $b= \ccp^2 + 2\ssp^2$, $c=\ssm^2 \ccm - \ssm \ssp \ccp$,
 and $d= \frac{1}{2}\left( 2 \ssm^2 + \ccp^2 + \ssp^2 \right)$.

\subsection{Polarization}
The polarization is defined as the average orientation per particle:
\begin{align}
\pp = \frac{\left< \nn P \right>}{\left< P \right>} =  \frac{\ssig}{2 \rho}.
\end{align}
Together with Eq. \eqref{app:sigma}, this gives
\begin{align}
p_x = - \frac{1}{4} \frac{1}{1 + \ssm^2 \fs^2}
    &\huge[
        \left( \ccp - 2 \ssm \ssp \fs \right) \nabR_x \fs
	\nonumber \\ &
        - \left( \ccp - \ssm \ssp \fs \right) \fs \nabR_x U 
    \huge],
\end{align}
and
\begin{align}
p_y =  \frac{1}{4} \frac{1}{1 + \ssm^2 \fs^2}
    &\huge[
        \left( 2 \ssp + \ssm \ccp \fs \right) \nabR_x \fs
	\nonumber \\ &
        - \left( \ssp + \ssm \ccp \fs \right) \fs \nabR_x U 
    \huge].
\end{align}

\subsection{Flux}\label{app:flux}
The flux in the y-direction (see Eqs. \eqref{app:J} and \eqref{app:sigma}) is
\begin{align}
J_y =&
    \fs A_{yj} \sigma_j - \frac{1}{2} B_{yx} \left(\nabR_x \fs \right) \rho
\nonumber \\
=&
   2 \fs A_{yj} p_j \rho - \frac{1}{2} B_{yx} \left(\nabR_x \fs \right) \rho
\nonumber \\
=&
    V_y \rho,
\end{align}
where
\begin{align}
V_y =&
    2 \fs A_{yx} p_x + 2 \fs A_{yy} p_y - \frac{1}{2} B_{yx} \left( \nabR_x \fs \right),
\nonumber \\
=&
    2 \fs  \left( \ssp p_x + \ccp p_y \right) - \frac{1}{2} \ssm \left( \nabR_x \fs \right).
\end{align}

\section{Torque}\label{torque_decomposition}
\begin{align}
\dt \rr = \FF(\rr) + \eeta.
\end{align}
The vector $\rr$ can be decomposed in a length and a orientation:
$\rr = r \nn$, with $|\nn| = 1$.
With Stratonovich integration one can then use $\nn {\cdot} d \nn=0$ to find the 
equations of motion for $\nn$ and $r$:
\begin{align}
\dt r = 
\nn {\cdot} \dt \rr =
\nn {\cdot} \FF + \nn {\cdot} \eeta,
\\
\dt \nn = 
    \frac{1}{r} \left(\I - \nn\nn\right) {\cdot} \FF
    +\frac{1}{r} \left(\I - \nn\nn\right) {\cdot} \eeta.
\end{align}
The Fokker-Planck equation corresponding to the last equation is
\begin{align}
\dt P(\nn,t) =&
    -\frac{1}{r} \left( n_x \partial_y - n_y \partial_x \right)
    \left[ \left( n_x F_y - n_y F_x \right) P \right]
	\nonumber \\ &
    + \frac{1}{r^2}  \left( n_x \partial_y - n_y \partial_x \right)^2 P,
\\
=&
- \frac{1}{r} \partial_{\theta} \left[ \left( n_x F_y - n_y F_x \right) P \right]
+ \frac{1}{r^2} \partial^2_{\theta} P,
\end{align}
where in the last step we used $\nn=( \cos(\theta), \sin(\theta) )$.
The SDE for $\nn$ is equivalent to
\begin{align}
\dt \nn &= 
    \frac{1}{r} \left(\I - \nn\nn\right) {\cdot} \FF
    +\frac{1}{r} \epsilon {\cdot} \nn \eta,
\end{align}
where
$ \epsilon=
\begin{bmatrix}
    0 & -1 \\
    1 & 0
\end{bmatrix} $,
and it is equivalent to
\begin{align}
 \dt \theta &= \frac{1}{r} n_x F_y - \frac{1}{r} n_y F_x +\frac{1}{r} \eta,\\
&=
    -\frac{1}{r} \nn{\cdot}\epsilon {\cdot} \FF +\frac{1}{r} \eta.
\end{align}

\section{Mapping to experimental dimer}\label{app:mapping_experimental_dimer}

Ref. \cite{ebbens2010selfassembled} reports experiments on dimers similar to the dimers in our work but with a costant swim force.
Example c of Ref. \cite{ebbens2010selfassembled} corresponds to $\phi_1 = \pi/2$ and $\phi_2 = \pi$, which is shown in Fig. 2(e) of the main text.

The dynamics of this dimer can be described by the following Langevin equations
\begin{align}
\dt \rr = \tilde{v} \tilde{\nn} + \sqrt{2 \tilde{D}} \xxi,
\label{eq:eomr_ebbens}
\\
\dt \theta = \tilde{\omega} + \sqrt{ 2 \tilde{D}_r} \eta,
\label{eq:eomTheta_ebbens}
\end{align}
see Eq. 1, 2 and 3 of Ref. \cite{ebbens2010selfassembled}.
The experimentally measured value of the parameters are
$\tilde{v} = 1.3~ \mu m/s$,
$\tilde{D} = 0.15~ \mu m^2 /s$,
$\tilde{\omega} = 1.1~s^{-1}$,
and $\tilde{D}_r = 1/16 ~s^{-1}$
(see Table I row c of Ref. \cite{ebbens2010selfassembled}).

The difference between the dimer in Fig. 2(d) and 2(e) of the main text
dissapears if there is no activity gradient.
The dynamics of the dimer in Fig. 2(e) with a constant activity are described by
\begin{align}
\dt x &= \frac{1}{2 \gamma} \fs (n_x - n_y) + \sqrt{\frac{T}{\gamma}} \xi_x,
\\
\dt y &= \frac{1}{2 \gamma} \fs (n_x + n_y) + \sqrt{\frac{T}{\gamma}} \xi_x,
\\
\dt \theta
  &= - \frac{1}{\gamma l} \fs + \sqrt{\frac{4 T}{\gamma l^2}} \eta,
\end{align}
where we have put the dimensions back in order to compare with the experimental system.

The $x$ component of the active force is $\frac{1}{2 \gamma} \fs (n_x - n_y)$,
which is equal to  $\frac{1}{\sqrt{2} \gamma} \fs \tilde{n}_x$,
and similarly, for the $y$ component of the active force is
 $\frac{1}{\sqrt{2} \gamma} \fs \tilde{n}_y$,
where $\tilde{n}_x$ and $\tilde{n}_y$ are the $x$ and $y$ components of 
the unit vector $\tilde{\nn}$ that points in the direction of the active force.
By comparing the previous equations with Eqs. \ref{eq:eomr_ebbens} and 
\ref{eq:eomTheta_ebbens} shows that
\begin{align}
\tilde{v} &= \frac{\fs}{\sqrt{2} \gamma},
\\
\tilde{D} &= \frac{T}{2 \gamma},
\\
\tilde{\omega} &= \frac{\fs}{\gamma l},
\\
\tilde{D}_r &= \frac{2 T}{\gamma l^2}.
\end{align}
The dimensionless swim force $ l \fs /(2 T)$ can becalculated in two ways:
\begin{align}
\frac{l}{2 T} \fs &= \frac{\tilde{v}^2}{2 \tilde{D} \tilde{\omega}} \approx 5.1,
\\
\frac{l}{2 T} \fs &= \frac{\tilde{v}}{\sqrt{2 \tilde{D} \tilde{D}_r}} \approx 9.5.
\end{align}
The two ways to calculate the dimensionless force do not agree because in our model we ignore hydrodynamics and the fact that connnecting the two ABPs has an effect on their activity.

The ratio of the density of the dimers in an active region and a region without activity
$\rho_{active}/\rho_{passive} \approx 4$, or
$\rho_{active}/\rho_{passive} \approx 16$,
depending on which way the force is estimated.

\bibliographystyle{apsrev}

\end{document}